\documentclass[journal,twocolumn]{IEEEtran}
\usepackage{}
\usepackage{graphicx}
\usepackage{epstopdf}
\usepackage{amssymb}
\usepackage{amsfonts}
\usepackage{amsmath}
\usepackage{algorithm}
\usepackage{algorithmic}
\usepackage{subeqnarray}
\usepackage{cases}
\usepackage{setspace}
\usepackage{subfigure,amsmath,amssymb,cite}

\usepackage{float}
\ifCLASSINFOpdf
\else
\fi
\begin{document}
	
	\title{Two Enhanced-rate Power Allocation Strategies for Active IRS-assisted Wireless Network}
	\author{Qiankun Cheng, Rongen Dong, Wenlong Cai, Ruiqi Liu, Feng Shu, and Jiangzhou Wang, \emph{Fellow IEEE}
		\thanks{This work was supported in part by the National Natural Science Foundation of China (Nos.U22A2002, and 62071234), the Hainan Province Science and Technology Special Fund (ZDKJ2021022), and the Scientific Research Fund Project of Hainan University under Grant KYQD(ZR)-21008, and the Collaborative Innovation Center of Information Technology, Hainan University (XTCX2022XXC07). \itshape(Corresponding author: Feng Shu).}
		\thanks{Qiankun Cheng and Rongen Dong are with the School of Information and Communication Engineering, Hainan University, Haikou 570228, China (e-mail: cqk1129@hainanu.edu.cn; dre2000@163.com).}
		\thanks{Feng Shu is with the School of Information and Communication Engineering, and Collaborative Innovation Center of Information Technology, Hainan University, Haikou 570228, China, and also with the School of Electronic and Optical Engineering, Nanjing University of Science and Technology, Nanjing 210094, China (e-mail: shufeng0101@163.com).}
		\thanks{Wenlong Cai is with National Key Laboratory of  Science and Technology on Aerospace Intelligence Control Beijing Aerospace Automatic Control Institute, Beijing, China.}
		\thanks{Ruiqi Liu is with the Wireless and Computing Research Institute, ZTE Corporation, Beijing 100029, China (e-mail: richie.leo@zte.com.cn).}
		\thanks{Jiangzhou Wang is with the School of Engineering, University of Kent, CT2 7NT Canterbury, U.K. (e-mail: j.z.wang@kent.ac.uk).}
	}
	\maketitle
	\begin{abstract}
		Due to its ability of overcoming the impact of double-fading effect, active intelligent reflecting surface (IRS) has attracted a lot of attention. Unlike passive IRS, active IRS should be supplied by power, thus adjusting power between base station (BS) and IRS having a direct impact on the system rate performance. In this paper, the active IRS-aided network under a total power constraint is modeled with an ability of adjusting power between BS and IRS. Given the transmit beamforming at BS and reflecting beamforming at IRS, the SNR expression is derived to be a function of power allocation (PA) factor, and the optimization of maximizing the SNR is given. Subsequently, two high-performance PA strategies, enhanced multiple random initialization Newton's (EMRIN) and Taylor polynomial approximation (TPA), are proposed. The former is to improve the rate performance of classic Netwon's method to avoid involving a local optimal point by using multiple random initializations. To reduce its high computational complexity, the latter provides a closed-form solution by making use of the first-order Taylor polynomial approximation to the original SNR function. Actually, using TPA, the original optimization problem is transformed into a problem of finding a root for a third-order polynomial.
		Simulation results are as follows: the first-order  TPA of SNR fit its exact expression well,  the proposed two PA methods performs much better than fixed PA in accordance with rate, and appoaches exhaustive search as the number of IRS reflecting elements goes to large-scale.
	\end{abstract}
	\section{Introduction}
	As new applications emerge for both consumers and vertical industries, there is an increasing demand for high-quality wireless communications which serving as a strong driving force for the evolution of wireless technologies \cite{g30,20,21,9410435}. On the other hand, high capacity wireless networks like the 5th generation (5G) and future 6th generation (6G) face several challenges in terms of costs and energy consumption. 
    Towards this background, it is broad consensus of both the academia and the industry to construct future networks with environment friendly components to reduce the total costs and energy consumption. As a low cost and low power consuming node, the intelligent reflecting surface (IRS) emerges as a strong candidate for future green wireless communications \cite{3,WCM_ris_standards,4}.
	
	In \cite{2}, by using deep reinforcement learning neural network, the transmit beamforming at BS and phase shifts at IRS can be simultaneously optimized to maximize the ergodic sum rate in an IRS-assisted multiuser downlink multiple-input single-output (MISO) system. In an IRS-aided MISO system \cite{3}, given the receiving end user's signal-to-noise (SNR) targets, two optimal solutions were presented to minimize the power consumed at base station (BS) by jointly optimizing the transmit beamforming of BS and phase shift matrix of IRS. In \cite{6}, it was showed that IRS can be used in multiple-input multiple-output (MIMO) system. The block coordinate descent method was proposed for maximizing the weighted sum rate by alternately optimizing precoding matrices in every BSs and phase shifts in IRS. Compared with the MIMO communication system without IRS, introducing IRS can significantly enhance the cell-edge user performance.
	
	Although passive IRSs, which consists of mainly passive and reflective elements, have significantly lower power consumption than active IRSs, recent studies demonstrated that active IRSs may have advantages under certain scenarios \cite{7,8,9,10}.
	Due to the multiplicative fading effect caused by the cascaded channel between the BS to the IRS and IRS to the user equipment \cite{9}, the power gain achieved by passive IRSs is limited in some scenarios. With active IRSs, which are implemented with power amplifiers, the impact of multiplicative fading can be mitigated \cite{8}. Since active IRSs possess more advanced capabilities than passive IRSs, to achieve globally optimized performance, it is intuitive to design the IRS and BS as a integrated system with direct controlling of the IRS by the BS. In this case, it is natural to assume that the active IRS will share a power source with the BS, which is also fair for performance comparison with other systems.
	
	In \cite{11,14}, the authors applied power allocation (PA) strategy into secure directional modulation (DM) system to improve the secrecy rate (SR)  performance. In \cite{12}, gradient descent method and Max-P-SINR-ANSNR method were proposed for maximizing the SR by optimizing the PA factor at transmitter in a secure spatial modulation system. In an active IRS-assisted DM network \cite{13}, two methods named Max-SR-FS and Max-SR-DG of maximizing SR were presented through alternately optimizing PA factors, the beamforming of BS, and phase shift matrix of active IRS under total power constraints. Thus, these works demonstrated that optimizing PA factors can greatly improve the performance of secure wireless communication systems.
	
	The above research works focused on the investigation of how to allocate power in secure wireless networks. To the best of our knowledge, there is no related research work concerning how to optimize PA between BS and IRS in an IRS-aided non-secure wireless network. Compared with traditional fixed PA, how many gains are achieved by a high-performance PA stragety in such a system. In this paper, we will concentrate our attention on developing two high-rate PA strageties to harvest the corresponding PA gains. Our main contributions are as follows:
	\begin{enumerate}
		\item First, an active IRS-assisted PA wireless network system model is established. And based on this model, the SNR expression as a function of PA factor is derived provided that the transmit and reflection beamforming vectors are designed well. To exploit the performance of traditional Newton's method, the enhanced multiple random initialization Newton's (EMRIN) method is proposed. Finally, a maximizer is used to find a better solution over conventional Newton's method. Simulation results show that the proposed EMRIN method can achieve a much better rate performance than Newton's method and outperforms fixed PA. As the random initialization time $K$ goes to a large value, the rate performance of the proposed EMRIN approach that of exhaustive search (ES).
		\item  However, the computational complexity of the above EMRIN is proportional to $K$. Apparently, for EMRIN, to approach the rate performance of ES means a high computational complexity. To reduce its computational complexity, a low-complexity closed-form solution, called Taylor polynomial approximation (TPA), is proposed. Here, a first-order Taylor polynomial is adopted to approximate the SNR function. Then, the corresponding problem of maximizing SNR is converted into a problem of finding a root of a third-order polynomial by using Cardan's formula. Simulation results show that: (a) the first-order SNR approximate function can fit the original SNR function well, (b) the proposed TPA may harvest substantial rate gains over fixed PA, (c) as the number of IRS elements tends to large-scale, the rate performance of the proposed TPA is close to those of ES and EMRIN.
	\end{enumerate}
	
	The reminder of this paper is organized as follows. In Section II, the system model with PA stragety is established, and two high-rate methods are proposed in Section III. Simulation results are presented in Section IV and conclusions are drawn in Section V.
	
	\textit{Notations}: During this paper, matrices and vectors are represented as uppercase letters and lowercase letters, respectively. $\mathbb{C}$ denotes a set of complex numbers. $(\cdot)^H$, $(\cdot)^T$, $(\cdot)^*$, $||\cdot||$, $\rm diag(\cdot)$, $||\cdot||_F$, $\mathbb{E}\{\cdot\}$ and $\mathfrak{R}\{\cdot\}$ denote the conjugate transpose, transpose, conjugate, Euclidean norm, diagonal, Frobenius norm, expectation and real part operations, respectively. $\mathbf{I}_{N}$ represents the $N \times N$ identity matrix.
	\section{system model}
	\begin{figure}[h]
		\centering
		\includegraphics[width=0.45\textwidth]{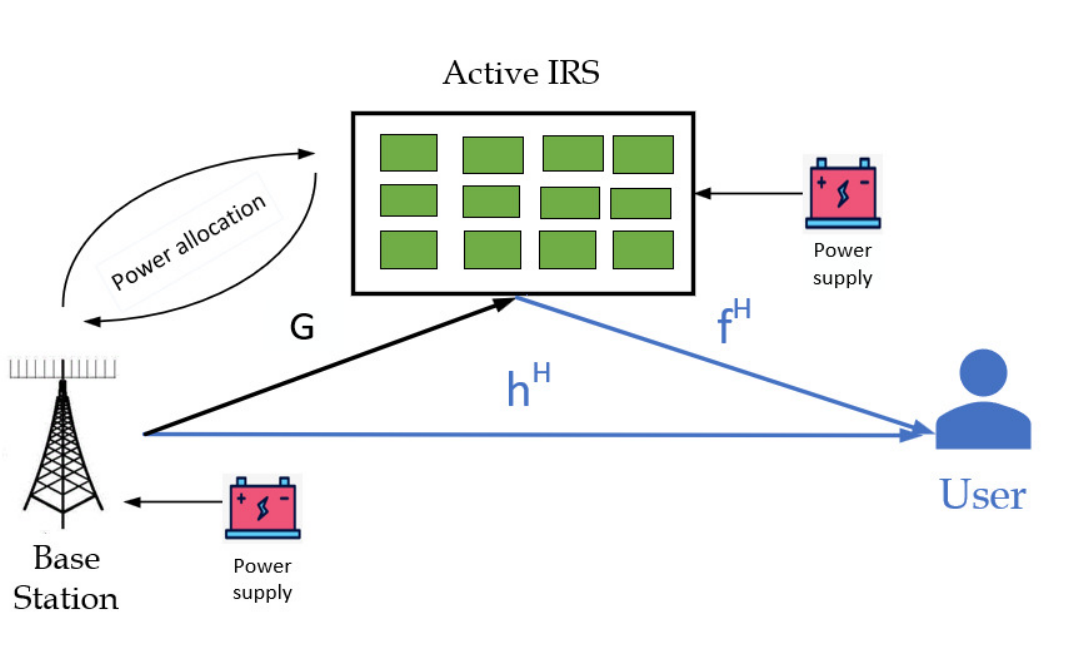}
		\caption{System model of an active IRS-assisted wireless network with PA}
	\end{figure}
	Fig.1 shows an active IRS-assisted wireless network with PA. BS is equipped with $M$ antennas. The user is equipped with single antenna, and active IRS is equipped with $N$ elements. $\mathbf{G}$ $\in$ $\mathbb{C}^{{N \times M}}$, $\mathbf{f}^H$ $\in$ $\mathbb{C}^{{1 \times N}}$ and $\mathbf{h}^H$ $\in$ $\mathbb{C}^{{1 \times M}}$ stand for the channels from BS to active IRS, active IRS to user, BS to user, respectively.
	
	The transmitted signal at BS is given by
	\begin{align}
		\mathbf{s}_B=\sqrt{\beta P_{max}}\mathbf{v}x,
	\end{align}
	where $\mathbf{v}$ $\in$ $\mathbb{C}^{{M \times 1}}$ is the transmit beamforming of BS that meets the condition $\mathbf{v}^H\mathbf{v}=1$, $P_{max}$ is the power sum of BS and active IRS, $\beta$ is the PA factor within interval [0,1], and $x$ denotes the transmit symbol satisfying $\mathbb{E}\left\{{|x|^2}\right\}=1$.
	
	For active IRS, let $a_m$ and $\theta_m$ represent amplification coefficient of the $m$-th element and phase shift of m-th element respectively, where ${m = 1, \cdots, N}$. $\mathbf{\Theta}=\operatorname{diag}\left(a_1e^{j\theta_1},\cdots,a_Ne^{j\theta_N}\right)$ denotes the reflective beamforming matrix of active IRS. The signal reflected by active IRS can written as:
	\begin{align}
		\mathbf{s}_I=\mathbf{\Theta}\mathbf{G}\mathbf{s}_B+\mathbf{\Theta}\mathbf{n}_I
		=\sqrt{\beta P_{max}}\mathbf{\Theta}\mathbf{G}\mathbf{v}x+\mathbf{\Theta}\mathbf{n}_I,
	\end{align}
	where $\mathbf{n}_I$ $\in$ $\mathbb{C}^{{N \times 1}}$ denotes the additive white Gaussian noise (AWGN) introduced by active IRS power amplifiers, $\mathbf{n}_I \sim \mathcal{CN}\left( {0,{\mathbf{{\sigma}}_I^2\mathbf{I}_{N}}}\right)$.
	
	Letting $\boldsymbol{\theta}=(a_1e^{j\theta_1},\cdots,a_Ne^{j\theta_N})^H$, the received signal at user is given by
	\begin{equation}
		\begin{aligned}
			y&=\sqrt{\beta P_{max}}(\mathbf{f}^H\mathbf{\Theta}\mathbf{G}+\mathbf{h}^H)\mathbf{v}x+\mathbf{f}^H\mathbf{\Theta}\mathbf{n}_I+z\\
			&=\sqrt{\beta P_{max}}(\rho\widetilde{\boldsymbol{\theta}}^H\operatorname{diag}(\mathbf{f}^H)\mathbf{G}+\mathbf{h}^H)\mathbf{v}x\\
			&+\rho\widetilde{\boldsymbol{\theta}}^H\operatorname{diag}(\mathbf{f}^H)\mathbf{n}_I+z,
		\end{aligned}
	\end{equation}
	where $\boldsymbol{\theta}=\rho\widetilde{\boldsymbol{\theta}}$, $\rho=\|\boldsymbol{\theta}\|_2$ and $\|\widetilde{\boldsymbol{\theta}}\|_2=1$. $z$ is the AWGN with distribution $z\sim\mathcal{C}\mathcal{N}(0,~\sigma^2_n)$.
	
	The power consumed at active IRS can be expressed as
	\begin{equation}
		\begin{aligned}
			P_{IRS}&= E\{\mathbf{s}_I^H\mathbf{s}_I\}=\beta P_{max}||\mathbf{\Theta}\mathbf{G}\mathbf{v}||_2^2+\sigma^2_I||\mathbf{\Theta}||_F^2\\
			&=\beta P_{max}\rho^2||\widetilde{\boldsymbol{\theta}}^H\operatorname{diag}(\mathbf{G}\mathbf{v})||_2^2+\sigma^2_I\rho^2=(1-\beta)P_{max},
		\end{aligned}
	\end{equation}
	with
	\begin{align}
		\rho
		=\sqrt{\frac{(1-\beta)P_{max}}{\beta P_{max}||\widetilde{\boldsymbol{\theta}}^H\operatorname{diag}(\mathbf{G}\mathbf{v})||_2^2+\sigma^2_I}}.
	\end{align}
	
	From (3), the signal-to-noise ratio (SNR) at user can be given by
	\begin{align}
		\text{SNR}
		&=\frac{\beta P_{max}||(\rho\widetilde{\boldsymbol{\theta}}^H\operatorname{diag}(\mathbf{f}^H)\mathbf{G}+\mathbf{h}^H)\mathbf{v}||_2^2}{\sigma^2_I\rho^2||\widetilde{\boldsymbol{\theta}}^H\operatorname{diag}(\mathbf{f}^H)||_2^2+\sigma^2_n}.
	\end{align}
	The problem of maximizing the SNR with respect to variables $\beta$, $\widetilde{\boldsymbol{\theta}}$, and $\mathbf{v}$ are cast as
	\begin{equation}
		\begin{aligned}
			\mathrm{(P1):}&\max_{\beta,\widetilde{\boldsymbol{\theta}},\mathbf{v}}~~~~~\text{SNR}
			=\frac{\beta P_{max}||(\rho\widetilde{\boldsymbol{\theta}}^H\operatorname{diag}(\mathbf{f}^H)\mathbf{G}+\mathbf{h}^H)\mathbf{v}||_2^2}{\sigma^2_I\rho^2||\widetilde{\boldsymbol{\theta}}^H\operatorname{diag}(\mathbf{f}^H)||_2^2+\sigma^2_n},\\
			&~~\text{s.t.}~~~~~~~0\le \beta \le 1,~ \widetilde{\boldsymbol{\theta}}^H\widetilde{\boldsymbol{\theta}}=1,~ \mathbf{v}^H\mathbf{v}=1,
		\end{aligned}
	\end{equation}
	given $\widetilde{\boldsymbol{\theta}}$ and $\mathbf{v}$ in advance, which reduces to
	\begin{equation}
		\begin{aligned}
			\mathrm{(P1-1):}&\max_{\beta}~~~~~		
			g(\beta)=\frac{a\beta^2+b\beta+2c\beta\sqrt{d\beta^2+e\beta+f}}{g\beta+h},\\
			&~~\text{s.t.}~~~~~~~~0 \le \beta \le 1,
		\end{aligned}
	\end{equation}
 where
	\begin{equation}
		\begin{aligned}
			a=&P_{max}^2||\mathbf{h}^H\mathbf{v}||^2||\widetilde{\boldsymbol{\theta}}^H\operatorname{diag}(\mathbf{G}\mathbf{v})||^2-\\
			&P_{max}^2||\widetilde{\boldsymbol{\theta}}^H\operatorname{diag}(\mathbf{f}^H)\mathbf{G}\mathbf{v}||^2,\\
			b=&P_{max}^2||\widetilde{\boldsymbol{\theta}}^H\operatorname{diag}(\mathbf{f}^H)\mathbf{G}\mathbf{v}||^2+P_{max}||\mathbf{h}^H\mathbf{v}||^2\sigma_I^2,\\
			c=&P_{max}\mathfrak{R}\{\widetilde{\boldsymbol{\theta}}^H\operatorname{diag}(\mathbf{f}^H)\mathbf{G}\mathbf{v}\mathbf{v}^H\mathbf{h}\},\\
			d=&-P_{max}^2||\widetilde{\boldsymbol{\theta}}^H\operatorname{diag}(\mathbf{G}\mathbf{v})||^2,\\
			e=&P_{max}^2||\widetilde{\boldsymbol{\theta}}^H\operatorname{diag}(\mathbf{G}\mathbf{v})||^2-\sigma_I^2P_{max}, f=P_{max}\sigma_I^2,\\
			g=&\sigma_n^2P_{max}||\widetilde{\boldsymbol{\theta}}^H\operatorname{diag}(\mathbf{G}\mathbf{v})||^2-\sigma_I^2P_{max}||\widetilde{\boldsymbol{\theta}}^H\operatorname{diag}(\mathbf{f}^H)||^2,\\
			h=&\sigma_I^2P_{max}||\widetilde{\boldsymbol{\theta}}^H\operatorname{diag}(\mathbf{f}^H)||^2+\sigma_n^2\sigma_I^2.
		\end{aligned}
	\end{equation}
	\section{Proposed Two High-performance PA Strategies}
	In this section, to enhance the achievable rate performance, two high-rate PA methods, EMRIN and TPA, are proposed for an active IRS-assisted wireless network, where the precoder at BS and the reflecting beamformer at IRS are fixed. The former is iterative while the latter is of closed-form. They make a substantial rate enhancement over conventional PA methods like Newton's method and fixed PA ones.
	
	\subsection{Proposed EMRIN Method}
	
	To improve the rate performance of the conventional Newton's method, multiple random initializations are combined with it, where the number of random initializations is denoted as $K$. In other words, for each random initialization, the Newton's method is used to iteratively compute the corresponding local optimal value. Eventually, a maximizer is adopted to find the maximum value of $K$ local optimal values.
	
	During the iteration process, the $k$-th random initial value is denoted as $\beta^{k,0}$ in the interval [0,1]. According to (8), the $i$-th iteration of the $k$-th initialization of the Newton's method is updated as follows
	\begin{align}\label{NetwonIExpre}
		\beta^{k,i}=\beta^{k,i-1}-\frac{1}{g''(\beta^{k,i-1})}g'(\beta^{k,i-1}),
	\end{align}
	where the first and second derivatives of function $g(\beta)$ are given as
	\begin{align}
		g'(\beta)= \frac{f_1(\beta)+f_2(\beta)}{(g\beta+h)^2},
	\end{align}
	\begin{align}
		g''(\beta)= \frac{(f_1'(\beta)+f_2'(\beta))(g\beta+h)-2g(f_1(\beta)+f_2(\beta))}{(g\beta+h)^3},
	\end{align}
	where
	\begin{align}
		f_1(\beta)=ag\beta^2+\frac{2cdg\beta^3+(ceg+2cdh)\beta^2+ceh\beta}{\sqrt{d\beta^2+e\beta+f}},
	\end{align}
	\begin{align}
		f_2(\beta)=2ah\beta+2ch\sqrt{d\beta^2+e\beta+f}+bh.
	\end{align}
	
	Via the Newton iterative expression in (\ref{NetwonIExpre}), the final solution for initialization time $k$ is $\beta^{k,o}$. Finally, we have the resulting solution
	\begin{align}
		\beta^{o}=\mathop{\rm argmax} \limits_{\beta \in S1} ~g(\beta),
	\end{align}
	where
	\begin{align}
		S1=\left\{\widetilde{\beta}^{(1,o)}, \cdots, \widetilde{\beta}^{(K,o)}\right\},
	\end{align}
	where
	\begin{equation}
		\begin{aligned}
			&\widetilde{\beta}^{(k,o)}=\begin{cases}
				{\beta}^{(k,o)}, {\beta}^{(k,o)} \in [0,1],\\
				~~~0, ~~{\beta}^{(k,o)}  \notin [0,1].
			\end{cases}
		\end{aligned}
	\end{equation}
	\begin{algorithm}
		\caption{Proposed EMRIN Method for Problem (P1-1)}
		\begin{algorithmic}[1]
			\STATE  Choose the proper random initialization time $K$, $k=0$, and convergence accuracy $\xi$.
			\REPEAT
			\STATE  Uniformly generate the random initial value $\beta^{k,0}$ in [0,1], $i=0$.
			\REPEAT
			\STATE Calculate $g'(\beta^{k,i})$ and $g''(\beta^{k,i})$ by (11) and (12).
			\STATE Update $\beta^{k,i+1}$ by (10).
			\STATE $i=i+1$.
			\UNTIL $|\beta^{k,i+1}-\beta^{k,i}|\le \xi$.
			\STATE Obtain $\widetilde{\beta}^{(k,o)}$ by (17).
			\STATE Using $\widetilde{\beta}^{(k,o)}$ to calculate $g(\widetilde{\beta}^{(k,o)})$ by (8).
			\STATE $k=k+1$.
			\UNTIL $k=K$.
			\STATE Obtain $\beta^{o}$ by (15).
		\end{algorithmic}
	\end{algorithm}
	This completes the derivation of the EMRIN method. Its detailed running process is listed in the above Table:  Algorithm 1. 
	\subsection{Proposed TPA Method}
	From the previous subsection, the proposed EMRIN method is iterative and its complexity is proportional to the number $K$ of randomization times. As $K$ grows up to a large value, its complexity will be high. To reduce its complexity, a low-complexity closed-form TPA will be presented as follows. Let us return to the objective function $g(\beta)$ in (P1-1)
	\begin{align}
		g(\beta)=\frac{a\beta^2+b\beta+2c\beta\sqrt{d\beta^2+e\beta+f}}{g\beta+h},
	\end{align}
	which can be rewritten as
	\begin{align}
		g(\beta)=\frac{a\beta^2+b\beta+2c\sqrt{f}\beta\sqrt{1+\frac{d}{f}\beta^2+\frac{e}{f}\beta}}{g\beta+h}.
	\end{align}
	Let us define
	\begin{align}
		g_1(\beta)=\sqrt{1+{\frac{d}{f}\beta^2+\frac{e}{f}\beta}}.
	\end{align}
	Considering $\beta \in [0,1]$, let us define
	\begin{align}
		\Delta x=\frac{d}{f}\beta^2+\frac{e}{f}\beta,
	\end{align}
	 then
	\begin{align}
		g_1(\beta)=g_1(\Delta x)=\sqrt{1+\Delta x}.
	\end{align}
	
	Applying the first-order Taylor formula expansion approximation to the above function $g_1(\beta)$, we have
	\begin{align}
		g_1(\Delta x)=\sqrt{1+\Delta x} \approx 1+\frac{1}{2}\Delta x,
	\end{align}
	which directly gives the result
	\begin{align}
		g_1(\beta)\approx 1+\frac{1}{2}\frac{d}{f}\beta^2+\frac{1}{2}\frac{e}{f}\beta.
	\end{align}
	Then, substituting the approximate function of $g_1(\beta)$ into (19), we can obtain the approximate function of $g(\beta)$ as follows
	\begin{equation}\label{ApproExpre}
		\begin{aligned}
			g(\beta) \approx g_2(\beta)=\frac{l\beta^3+m\beta^2+n\beta}{g\beta+h},
		\end{aligned}
	\end{equation}
	where
	\begin{align}
		l=\frac{cd\sqrt{f}}{f}, m=a+\frac{ce\sqrt{f}}{f}, n=b+2c\sqrt{f}.
	\end{align}
	
	To find its maximum value in the interval [0,1], taking the first derivative of its approximate function $g_2(\beta)$ in (25) with repect to $\beta$ equal zero yields
	\begin{align}\label{AR-FD}
		g_2'(\beta)=\frac{p\beta^3+q\beta^2+r\beta+s}{(g\beta+h)^2}=0,
	\end{align}
	where
	\begin{align}
		p=2gl, q=mg+3lh, r=2mh, s=nh.
	\end{align}
	The equation (\ref{AR-FD}) reduces to
	\begin{align}
		p\beta^3+q\beta^2+r\beta+s=0.
	\end{align}
	By using the Cardano's formula, three candidate solutions to the above equation are represented as follows
	\begin{equation}
		\begin{aligned}
			&\beta_1=\gamma+\sqrt[3]{-u+\eta}+\sqrt[3]{-u-\eta},\\
			&\beta_2=\gamma+\omega\sqrt[3]{-u+\eta}+\omega^2\sqrt[3]{-u-\eta},\\
			&\beta_3=\gamma+\omega^2\sqrt[3]{-u+\eta}+\omega\sqrt[3]{-u-\eta}.
		\end{aligned}
	\end{equation}
	where $\gamma=-q/3p$, and $\eta=(u^2+t^3)^{1/2}$ with
	\begin{align}
	t=\frac{3pr-q^2}{9p^2}, u=\frac{27p^2s-9pqr+2q^3}{54p^3},\omega=\frac{-1+\sqrt{3}\rm i}{2}.
	\end{align}
	
	Considering $\beta \in [0,1]$, we need to judge whether $\beta_1$, $\beta_2$ and $\beta_3$ are within the interval [0,1], then the new forms of three-candidate solutions are constructed as follows
	\begin{equation}
		\begin{aligned}
			&\widetilde{\beta_i}=\begin{cases}
				\beta_i, \beta_i \in [0,1],\\
				0, ~\beta_i \notin [0,1].
			\end{cases}
		\end{aligned}
	\end{equation}
	where $i={\left\{1,2,3\right\}}$, the optimal solution of $\beta$ is
	\begin{align}
		\beta^{o}=\mathop{\rm argmax} \limits_{\beta \in S2} ~g(\beta),
	\end{align}
	where set $S2$ is defined as
	\begin{align}
		S2=\left\{0, \widetilde{\beta}_1, \widetilde{\beta}_2, \widetilde{\beta}_3, 1\right\}
	\end{align}
	in accordance with the Extreme Value theorem in \cite{Burden}.
	This completes the derivation of the proposed TPA method.
	\subsection{Complexity analysis}
	 The computational complexities of the proposed EMRIN, TPA, and ES are  $\mathcal{O}\left\{KI\right\}$ float-point operations (FLOPs), $\mathcal{O}\left\{M\right\}$ FLOPs, and $\mathcal{O}\left\{J\right\}$ FLOPs, respectively. Here $I$ denotes the number of iterations per initialization of EMRIN method, $M$ the number of elements in set $S2$, and $J$ the number of subintervals of ES. In the following simulation, to achieve their best performance,  the values of $M$, $K$, and $J$  are taken to be 5, 256, and 10001. In summary, their compexities are as follows: TPA$\ll $EMRIN$\ll$ES. 
	\section{Simulation Results}
	In this section, simulation results are presented with parameter setup as follows. The locations of BS, IRS, and user are set to (0 m, 0 m, 0 m), (100 m, 0 m, 10 m), and (50 m, 30 m, 0 m), respectively. Following \cite{6}, the randomly generated channel matrix $\mathbf{G}$, channel vectors $\mathbf{f}$ and $\mathbf{h}$ follow the Rayleigh distribution, channel path fading factors from BS to IRS, IRS to user, and BS to user are set to 2.1, 2.1, 4.0, respectively. The number of BS antennas is chosen as follows:  $M=2$, noise power $\sigma_I^2 = \sigma^2_n = -100$ dBm.
	\begin{figure*}
		\setlength{\abovecaptionskip}{-5pt}
		\setlength{\belowcaptionskip}{-10pt}
		\centering
		\begin{minipage}[t]{0.33\linewidth}
			\centering
			\includegraphics[width=2.56in]{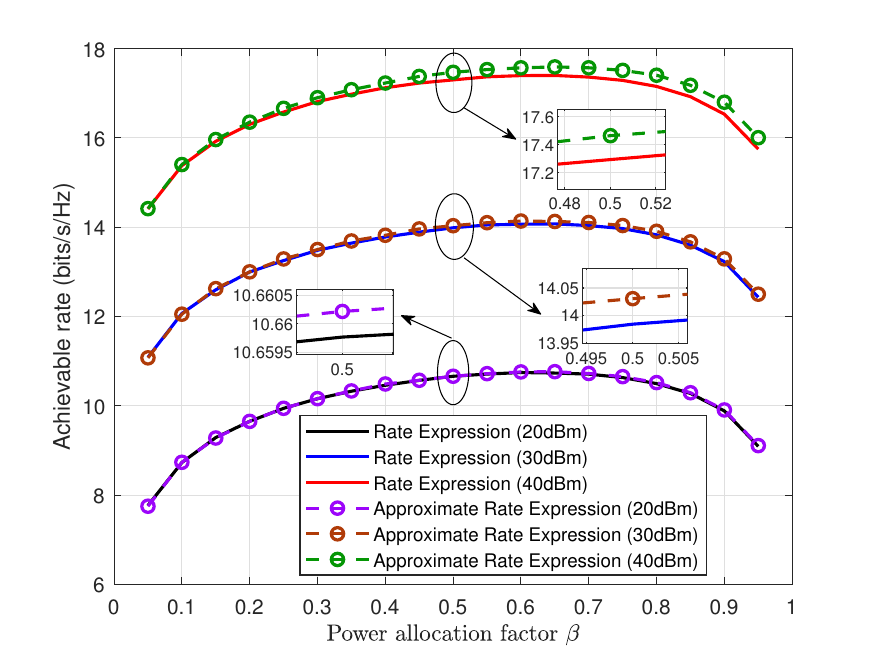}
			\caption{Achievable rate versus PA factor $\beta$}\label{Rate-to-Beta}
		\end{minipage}%
		\begin{minipage}[t]{0.33\linewidth}
			\centering
			\includegraphics[width=2.56in]{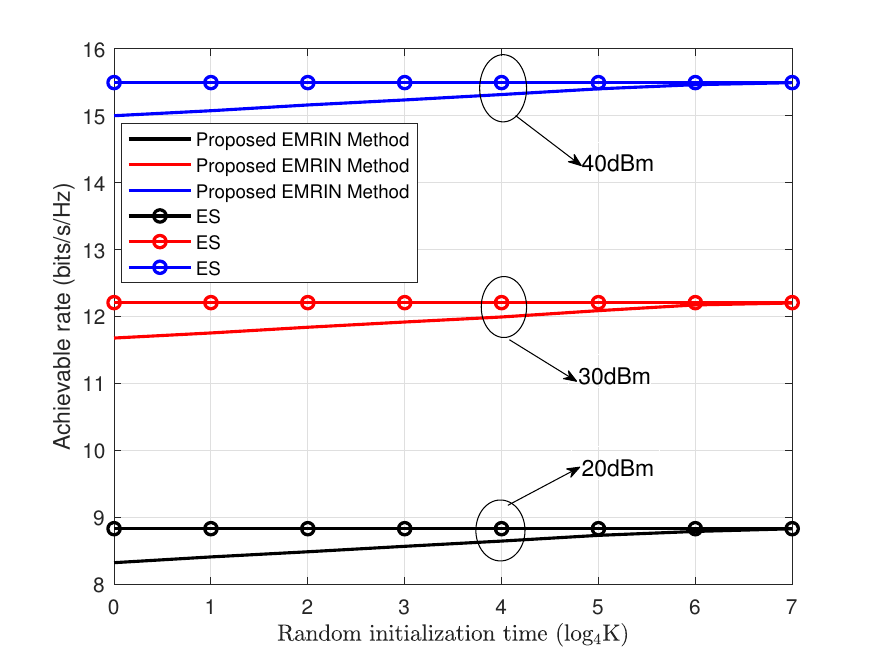}
				\caption{Achievable rate versus random initialization time $K$}\label{Rate-to-RIT}
		\end{minipage}
		\begin{minipage}[t]{0.33\linewidth}
			\centering
			\includegraphics[width=2.56in]{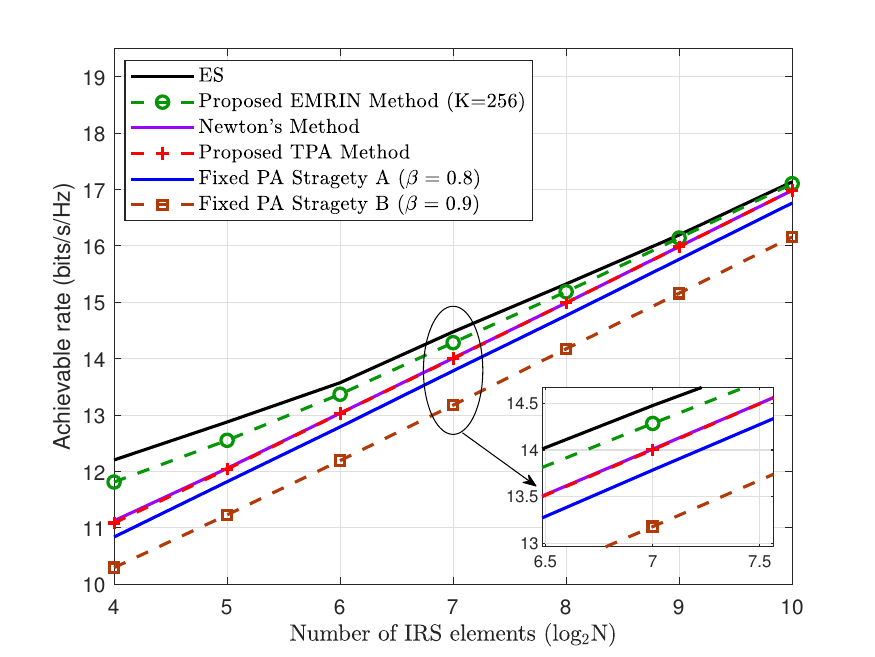}
			\caption{Achievable rate versus number of IRS elements $N$}\label{Rate-to-N}
		\end{minipage}
	\end{figure*}

	
	Fig.~\ref{Rate-to-Beta} illustrates the curves of rate expression and its approximate expression versus the PA factor $\beta$ for three distinct power sums: 20 dBm, 30 dBm, and 40 dBm with $N=128$. From Fig.~\ref{Rate-to-Beta}, it is seen that the approximate rate expression in (\ref{ApproExpre}) can fit the exact rate one well.
	
	
	Fig.~\ref{Rate-to-RIT} demonstrates the achievable rate of the EMRIN method versus its random initialization time $K$, where $N=16$, $P_{max}=20$ dBm, $30$ dBm, and $40$ dBm, respectively. From Fig.~\ref{Rate-to-RIT}, it is seen that as the number of random initialization times increases, the performance of EMRIN method gradually approaches the optimal ES. Considering the rate loss is lower than 0.2 bit for $K\ge256$, the number $K$ of random initializations is chosen to be 256.
	
	Fig. \ref{Rate-to-N} plots the achievable rates of the proposed EMRIN and TPA methods versus the number of IRS elements $N$ with fixed PA stragety A $(\beta=0.8)$, fixed PA strategy B $(\beta=0.9)$, Newton's method and ES as performance benchmarks. Here, $P_{max} = 30$ dBm. From Fig. 4, it is seen that the proposed two methods can achieve an obvious rate performance gains over Newton's method, and fixed PA strategies A and B. As $N$ goes to large-scale, they are close to ES in terms of rate.  Especially, the EMRIN method can provide more than 1 bit rate gain over fixed PA strageties A and B as $N$ tends to small-scale. The rate performance of the proposed TPA method is slightly better than that of the traditional Newton's method. In summary, six methods have a  increasing order in rate performance as follows: ES, EMRIN, TPA, Newton, fixed PA strategy A, and fixed PA strategy B. Considering the closed-form expression of the proposed TPA, it strikes a good balance between complexity and performance.
	\section{Conclusion}
	In this paper, we have focused our attention on the investigation of adjusting the PA between BS and IRS in an active IRS-assisted wireless network under the condition that the beamforming vectors at BS and IRS are given. First, the SNR expression has been derived as a function of PA factor and the corresponding general problem of maximizing SNR was also formulated. To enhance the rate performance of Newton's method, the EMRIN method was designed. To reduce its high complexity, the closed-form TPA was proposed to achieve an extremely low-complexity by making use of the first-order Taylor polynomial approximation. Simulation results confirmed that the proposed EMRIN and TPA obviously outperform existing methods fixed PA and Newton. As the number of IRS elements tends to large-scale, their rate performance is close to that of the optimal ES.
	\ifCLASSOPTIONcaptionsoff
	\newpage
	\fi
	\bibliographystyle{IEEEtran}
	\bibliography{IEEEfull,reference}
\end{document}